\documentclass[a4paper,11pt]{article}
\pdfoutput=1 

\usepackage{jinstpub} 

\usepackage{lineno}

\usepackage{multirow}

\title{\boldmath Optimized scintillation strip design for the DANSS upgrade}

\author[a,b]{I.~Alekseev,}
\author[c]{V.~Belov,}
\author[c,d]{A.~Bystryakov,}
\author[b]{M.~Danilov,}
\author[a,e]{A.~Ershova,}
\author[c]{D.~Filosofov,}
\author[c]{M.~Fomina,}
\author[c,f]{S.~Kazartsev,}
\author[a,e]{A.~Kobyakin,}
\author[g]{N.~Kozlenko,}
\author[c]{A.~Kuznetsov,}
\author[h]{I.~Machikhiliyan,}
\author[c,i]{F.~Mamedov,}
\author[c]{D.~Medvedev,}
\author[a]{V.~Nesterov,}
\author[g]{D.~Novinsky,}
\author[a,j]{K.~Perminov,}
\author[c]{I.~Rozova,}
\author[c]{N.~Rumyantseva,}
\author[a]{V.~Rusinov,}
\author[c]{A.~Salamatin,}
\author[a]{E.~Samigullin,}
\author[c]{Ye.~Shevchik,}
\author[c]{M.~Shirchenko,}
\author[c]{Yu.~Shitov,}
\author[a,b]{N.~Skrobova,}
\author[a]{A.~Starostin,}
\author[i]{I.~Štekl,}
\author[a,b,e,1]{D.~Svirida\note{Corresponding author.},}
\author[a]{E.~Tarkovsky,}
\author[e]{A.~Yakovleva,}
\author[c]{E.~Yakushev,}
\author[c]{I.~Zhitnikov}
\author[c]{and D.~Zinatulina}

\affiliation[a]{Alikhanov Institute for Theoretical and Experimental Physic NRC "Kurchatov Institute",\\ 
    B. Cheremushkinskaya str. 25, Moscow, 117218, Russia}
\affiliation[b]{Lebedev Physical Institute of the Russian Academy of Sciences, \\ Leninskiy avenue 53, Moscow, 119991, Russia}
\affiliation[c]{Joint Institute for Nuclear Research, \\ Joliot-Curie str. 6, Dubna, Moscow region, 141980, Russia}
\affiliation[d]{Dubna State University, \\ Universitetskaya str. 19, Dubna, Moscow Region, 141982, Russia}
\affiliation[e]{Moscow Institute of Physics and Technology, \\ Institutskiy lane 9, Dolgoprudny, Moscow Region, 141701, Russia}
\affiliation[f]{Voronezh State University, \\ Universitetskaya square 1, Voronezh, 1394018, Russia}
\affiliation[g]{Konstantinov Petersburg Nuclear Physics Institute NRC "Kurchatov Institute", \\
    Orlova roshcha 1, Gatchina, Leningrad Region, 188300, Russia}
\affiliation[h]{Federal State Unitary Enterprise Dukhov Automatics Research Institute, \\ Sushchevskaya str. 22, Moscow 127055, Russia}
\affiliation[i]{Institute of Experimental and Applied Physics, Czech Technical University in Prague, \\
    Horska 3a/22, CZ 12800 Prague 2, Czech Republic}
\affiliation[j]{New Economic School, \\ Nobelya str. 3, Skolkovo Innovation Center, Moscow, Russia}

\emailAdd{Dmitry.Svirida@itep.ru}

\abstract{DANSS is a one cubic meter plastic scintillator detector with a primary goal of sterile neutrino searches at 
a commercial nuclear reactor. Due to its highly advantageous location, fine segmentation and ability to change the distance to
the neutrino production origin, DANSS is ahead of many similar experiments around 
the world in terms of the counting rate, signal to background ratio and sterile neutrino exclusion regions. 
Yet a moderate energy resolution of the detector prevents further progress in the physics program. The main challenge
of the planned upgrade is to achieve an energy resolution of 12\% at 1~MeV. The new design of the main sensitive element -- 
the plastic scintillation strip -- is the most important step forward. The strip prototypes were manufactured and tested at 
the pion beam of the PNPI synchrocyclotron. More than twice higher light output together with fairly flat detector response
uniformity, longitudinal timing information and other optimizations will help to reach the upgrade goal. This paper
discusses the drawbacks of the current strip version, outlines the new features of the proposed upgrade, describes the beam
test procedure and presents the test results reflecting the advantages of the new strip design in comparison with the current
version.  
}


\begin{document}
\maketitle
\flushbottom

\section{Introduction}

\begin{figure}[htbp]
\centering
\begin{tabular}{ccc}
\includegraphics[width=.40\textwidth, trim= 0mm 0mm 0mm 0mm, clip]{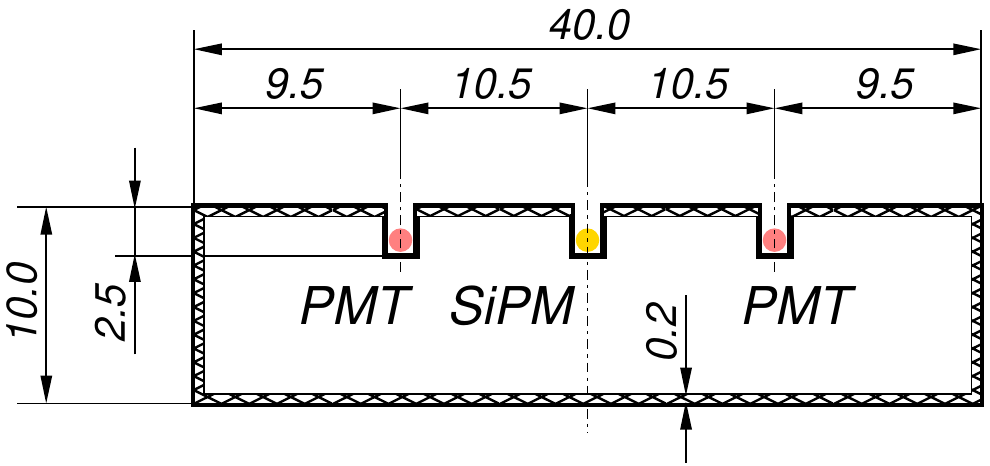} & &
\includegraphics[width=.505\textwidth, trim= 0mm 0mm 0mm 0mm, clip]{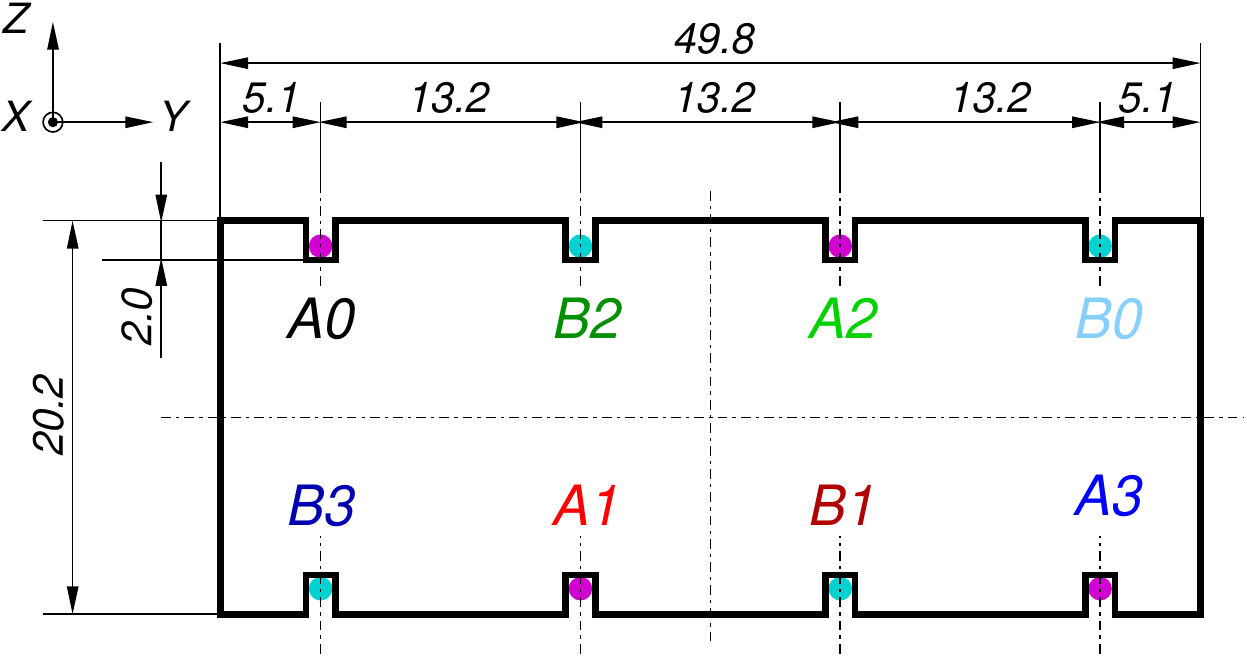} \\
a) & & b) \\
\end{tabular}
\caption{\label{fig:strips} Current (a) and future (b) design of the scintillation strips for the DANSS detector (transverse section).
The groove width is 1.5~mm, the fiber diameter -- 1.2~mm, in both cases.}
\end{figure}

DANSS is a reactor antineutrino detector~\cite{danss} composed of 2500 one meter long plastic scintillator strips. The 
transverse section view of a strip is given in figure~\ref{fig:strips}a. 25 parallel strips make a 1~cm thick layer, 
and 100 layers form a 1~cubic meter sensitive volume. The strips in adjacent layers are laid in perpendicular directions.
The surface 200~$\mu$m layer of each strip contains titanium oxide for light reflection and gadolinium
oxide for neutron capture. The light collection is done by means of three wavelength shifting (WLS) fibers embedded 
into longitudinal grooves filled with optical gel~\cite{gel}. The central fiber is read out by a silicon photomultiplier (SiPM),
individually for each strip. Side fibers of every 5 strips from 10 layers with parallel strip direction are combined into 
bundles of 100 on cathodes of conventional PMTs. Both SiPM and PMT readout happens at the same side of the strip. 
Fiber edges at the opposite side of the strip are polished and covered with a silver paint~\cite{mirror}
to achieve a mirror reflection.

Because the detector contains no flammable or other dangerous materials it is allowed to be placed
in the direct proximity of a commercial nuclear power reactor. The distance from the sensitive volume to the 
reactor core ranges from 10.9~m to 12.9~m center to center. The detector elevation is changed with 
the help of a lifting platform, typically three times a week. The core itself and the surrounding reactor 
infrastructure are located above the detector and provide significant shielding of about 50 m.w.e. from 
cosmic radiation. Due to so many advantages the DANSS detector sets world records in neutrino detection
at very short baselines. The counting rate reaches 5000 antineutrino events per day, while the total accumulated
statistics already exceeds 5~million antineutrinos after less than 5~years of data taking. An additional advantage
arises from the fine segmentation of the detector which provides very clean identification of the inverse beta
decay (IBD) events produced by antineutrinos. The whole configuration of passive and active shielding together
with the fine segmentation of the sensitive volume allows reaching an excellent signal to background ratio above 50.

At the same time, the DANSS detector has one essential drawback: quite moderate energy resolution
of 34\% at 1 MeV. Both stochastic and constant terms make significant contributions. The average light
yield leaves much to be desired and quantitatively is only 18.9 photoelectrons/MeV for 
SiPMs and 15.3 p.e./MeV for PMTs (about 34 p.e./MeV in total).
The constant term arises from inhomogeneities of various types, of which the light yield nonuniformity plays the dominant role. 
The correction for the longitudinal attenuation can only be made for the events when two or more adjacent perpendicular
strips are hit and the coordinate along the strips can be determined. Thus most of the low energy hits with the
energy deposit in only one strip loose the attenuation correction which reaches 30\%/m. The transverse hit coordinate
can not be measured or otherwise corrected for under the conditions of the running detector. Yet dedicated 
studies~\cite{lyield} give an estimate of at least 8\% as a contribution to the energy resolution from the transverse light 
yield nonuniformity. Also, the reflective surface layer of the strips contains a considerable amount of heavy titanium which 
efficiently captures $\gamma$-quants with no light emission. The situation is made even worse by the extrusion technology of the strips, which only 
poorly maintains this layer thickness that spreads from 100 to 300 micrometers. Beyond that the grooves
themselves produce additional insensitive areas and deteriorate the spatial homogeneity. This is inevitable with the WLS fiber
technology readout, but the groove depth can also be a subject of optimization. 

\section{New strip design}

The energy resolution is the main limiting factor for the further progress of the DANSS experiment. That is why the detector
upgrade is intended in the nearest future with the primary goal of the improvement of the detector energy resolution. The second,
but not the least important challenge is to increase the detector sensitive volume preserving the existing infrastructure of the
bearing frame and the lifting platform. Due to the limited funding of the upgrade the number of the digitization channels has
to be kept at the same level. 

The new strip design (see figure~\ref{fig:strips}b) takes into account many of the drawbacks specific to the current version
and pays attention to the mentioned challenges and limitations. The strip is now twice as thick and 20\% wider but has 8 grooves
for WLS fibers distributed on both top and bottom surfaces. The groove depth is decreased to a practical minimum, while the groove
positions are optimized according to the simulation (see the next section). The fibers are read out by SiPMs from both strip sides, 
4 SiPMs at each side, in an alternating manner. In figure~\ref{fig:strips}b the fibers that are coupled to the SiPMs at the side~A
are shown in magenta and denoted as \emph{An}. Those coupled to the side~B are colored in cyan and marked as \emph{Bn}. 
The colors of the labels approximately resemble the colors of the plots in the figures below. Due to the double-sided readout the timing
information can be used for the longitudinal coordinate estimates even in case of isolated hits. The strips are cut out and then
machined from a block of a bulk polystyrene at IPTP (Dubna, Russia)~\cite{iftp}. This technology of bulk polymerization is known to 
result in better scintillation light yield compared to the extrusion process. The light reflection layer on the strip surface is 
created by a chemical foaming of the polystyrene itself, with no admixtures of other chemicals. The process is thoroughly 
worked out at UNIPLAST (Vladimir, Russia)~\cite{uniplast}
and provides a much better control on the layer thickness, which is maintained at the level of 100 micrometers. The necessary 
amount of gadolinium for the IBD neutron capture is added separately from the scintillation strips in the form of the oxide powder
mixed into a polyethylene film, which is also produced at~\cite{uniplast}. The film is put between the strip layers, while its
thickness is well preserved at 250 micrometers. The removal of the vacuum PMTs frees space for a larger sensitive volume. 
The new strips are 120~cm long and comprise the scintillation cube with this side length by forming 60 layers of 24 strips
each. Each successive layer, as earlier, is laid perpendicularly to the previous one. The four SiPMs at each strip side will 
be wired together in the final design to form a single digitization channel. Nevertheless each SiPM will have an individually
tunable power supply in order to equalize single pixel responses. During the calibration process three of four SiPMs 
at a given side will be switched off except the one currently under adjustment. The total number of readout channels will 
thus be 2880, and the increase is only 15\%.  At the same time, the sensitive volume grows by more than 70\%. Yet, for this 
study, all the SiPMs at the strips under tests were read out and digitized individually.

\section{Simulations}

\begin{figure}[htbp]
\centering
\begin{tabular}{cc}
\includegraphics[width=.45\textwidth]{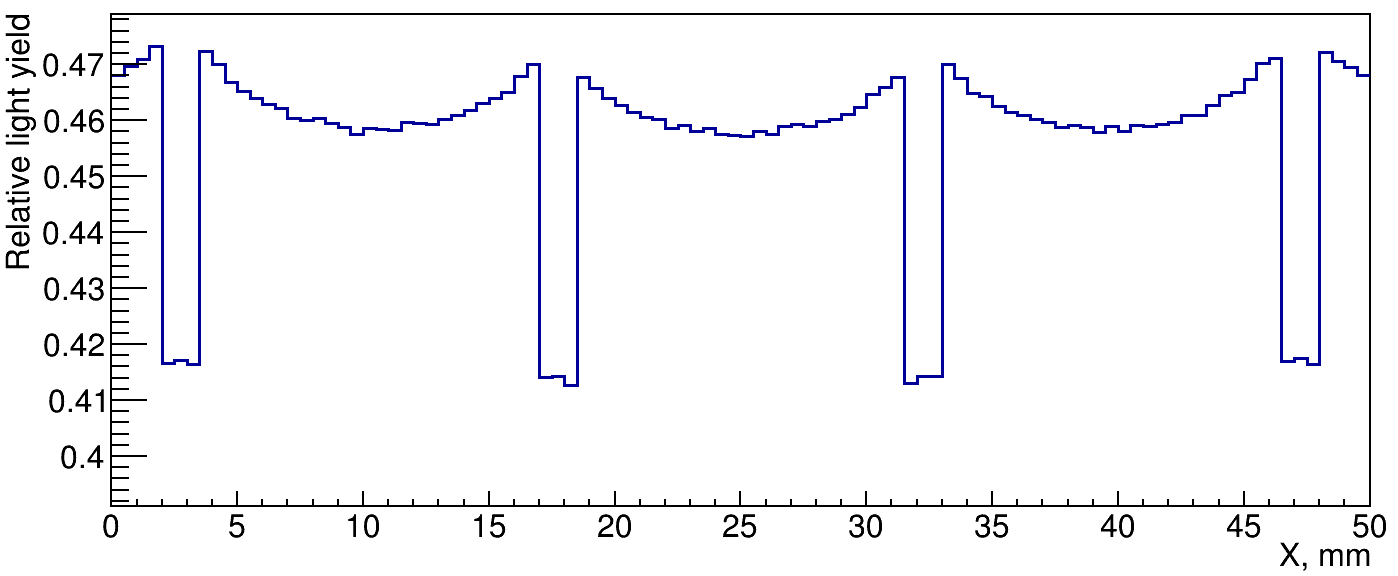} & 
	\multirow{4}{*}[.12\textwidth]{
	    \begin{tabular}{c}
	    \includegraphics[width=.45\textwidth]{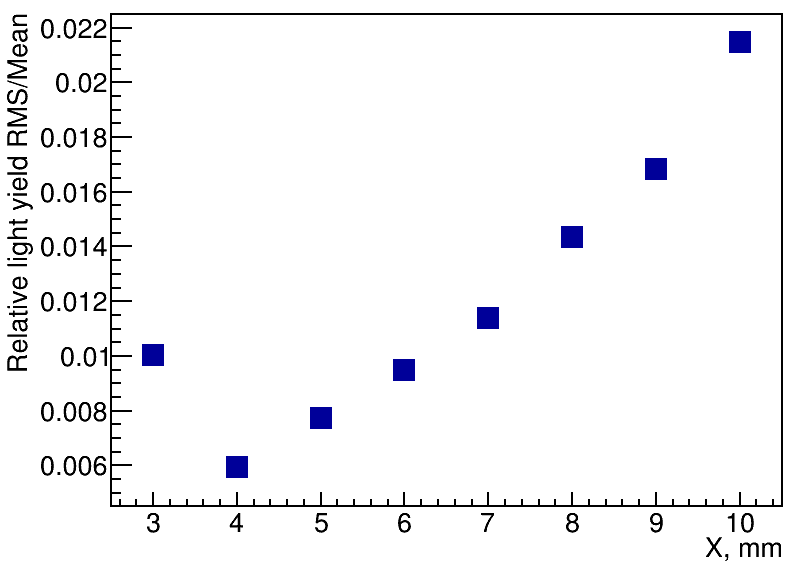} \\  
	    c) \\
	    \end{tabular} 
	} \\
a) & \\
\includegraphics[width=.45\textwidth]{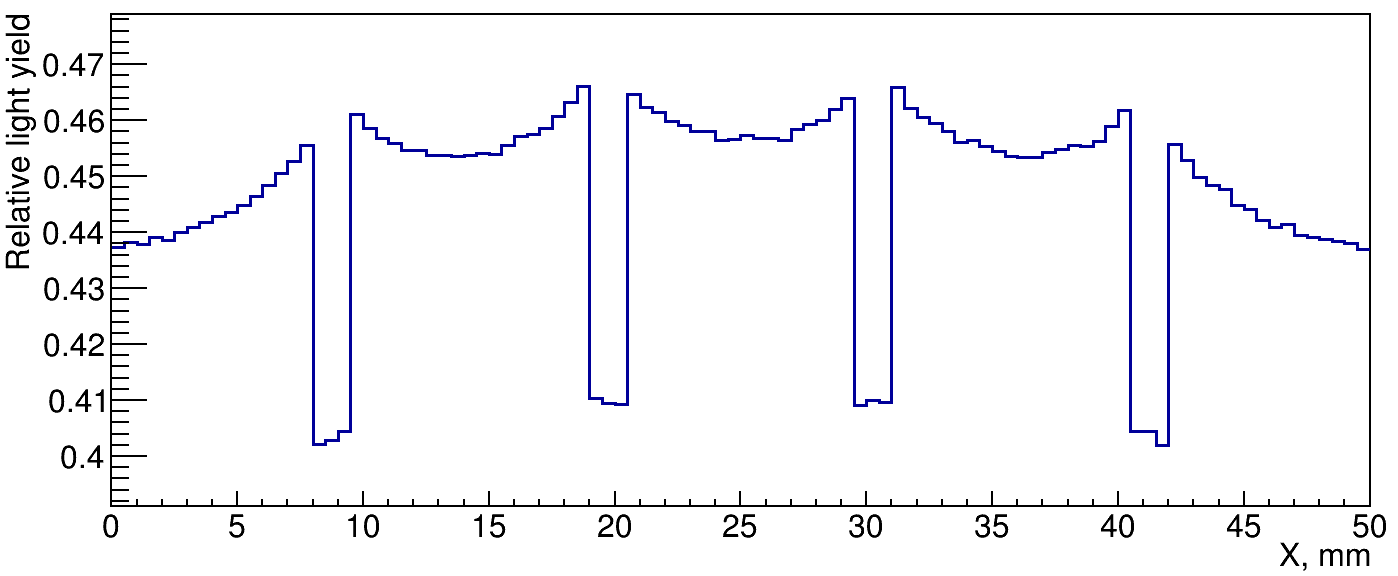} & \\
b) & \\
\end{tabular}
\caption{\label{fig:mc} Simulated light yield as a function of the perpendicular track position for the outer
grooves at 3~mm (a) and 9~mm (b) from the strip sides; (c) -- light yield RMS, normalized by its mean value, vs outer groove position. 
}
\end{figure}

A simple toy Monte-Carlo simulation was performed to estimate the influence of the groove positions on the homogeneity of the light collection.
Photons were randomly emitted from track segments perpendicular to the strip plane, and traced until they got captured by one of the fibers 
or disappeared on surfaces with finite diffuse reflection capabilities. Absorbtion in the scintillator, optical gel and fibers was also 
taken into account. Only the segment parts in the scintillator volume emitted the light 
thus accounting for shorter track lengths at the location of the grooves. The grooves with fibers were filled with optical gel and covered
with a fully reflecting foil. The track segments were sequentially positioned with 1~mm step from the shortest side of the strip and the 
photon capture efficiency was estimated for each position. Given that the number of photons emitted from the shorter tracks was 
proportionally smaller, the calculated values are fully equivalent to the relative light yield from the perpendicular tracks. The numbers
are normalized in such a way that for a full 20~mm track they represent the probability of photon capture by any of the WLS fibers.
Such procedure was repeated for several strip geometries with the outer grooves located 3 to 10~mm from the strip sides and the central
grooves spaced equally. Two examples of the efficiency distributions are presented in figures~\ref{fig:mc}a and \ref{fig:mc}b.  
The light collection from the very sides of the strip becomes insufficient when the grooves are moved closer to
the strip center, while larger distance between the grooves leads to a more pronounced inhomogeneity in the areas between the grooves
and some excess of the light at the sides of the strip. Figure~\ref{fig:mc}c shows how the spread of the light yield depends on the
distance of the outer grooves from the strip side. The optimal position is at about 4~mm from the side and this geometry
is much better than the case of equal spacing of 10~mm.

The optimal groove spacing of 4~mm from the strip side seems impractical, leaving too little of the flat strip surface near the side.
Yet 5~mm position combines reasonable simulated homogeneity with sufficient distance from the side and this option was chosen for further
tests.

\section{Test conditions}

Seven strips were manufactured according to the chosen design and placed into a lightproof box together with the front-end 
electronics. The 1.2~mm WLS fibers used in the test strips were of the same brand as those in the current DANSS detector: 
Y-11(200)M from Kuraray~\cite{kura}. The tests were performed at the pion beam 1 of the synchrocyclotron 
SC-1000~\cite{sc1000} at PNPI of NRC KI.

\begin{figure}[htbp]
\centering
\includegraphics[width=.85\textwidth]{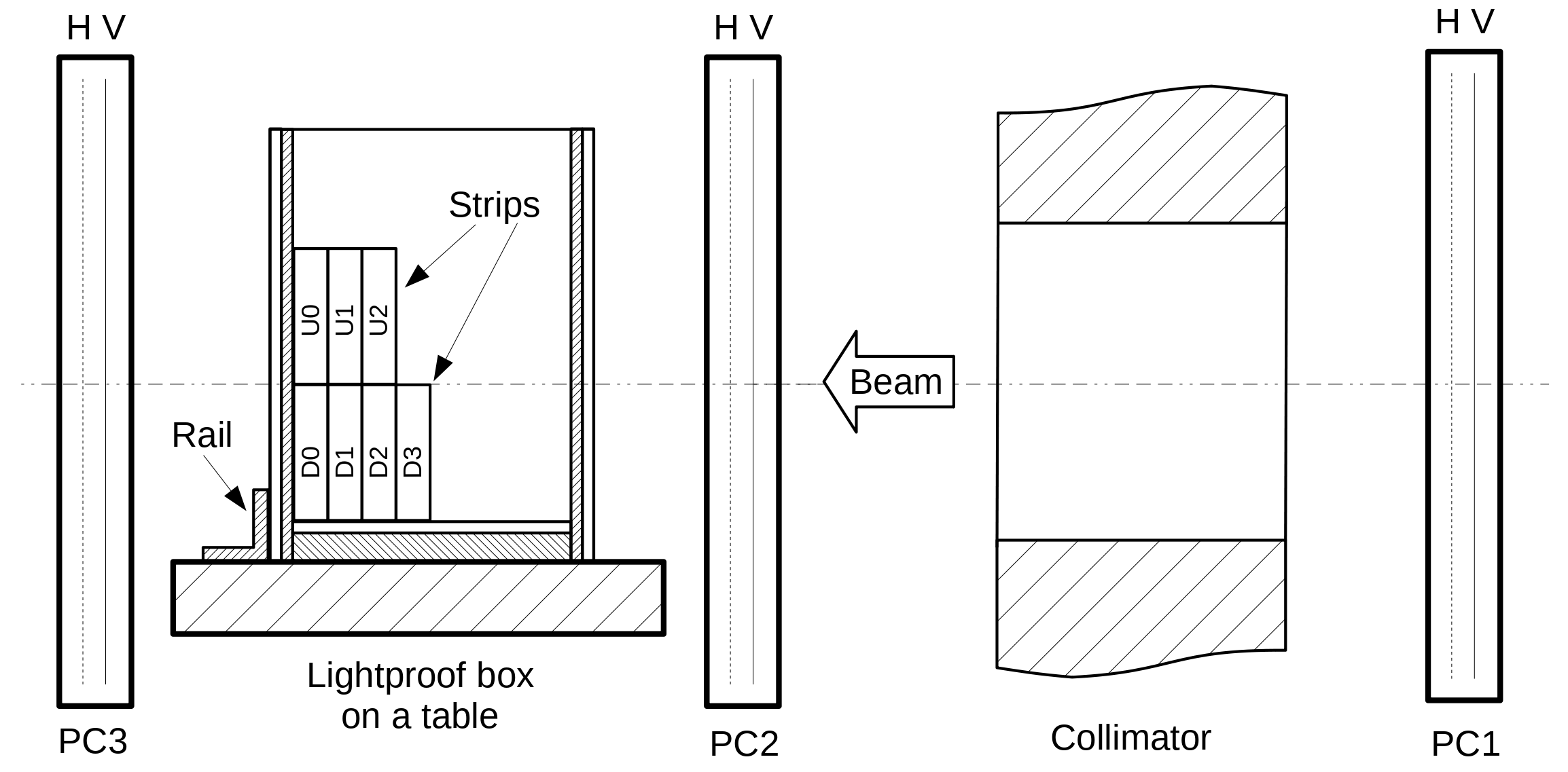}  
\caption{\label{fig:setup} 
Test setup at the pion beam 1 of the PNPI synchrocyclotron.}
\end{figure}

\begin{figure}[htbp]
\centering
\begin{tabular}{ccc}
\includegraphics[width=.32\textwidth, trim= 5mm 0mm 16mm .08\textwidth, clip]{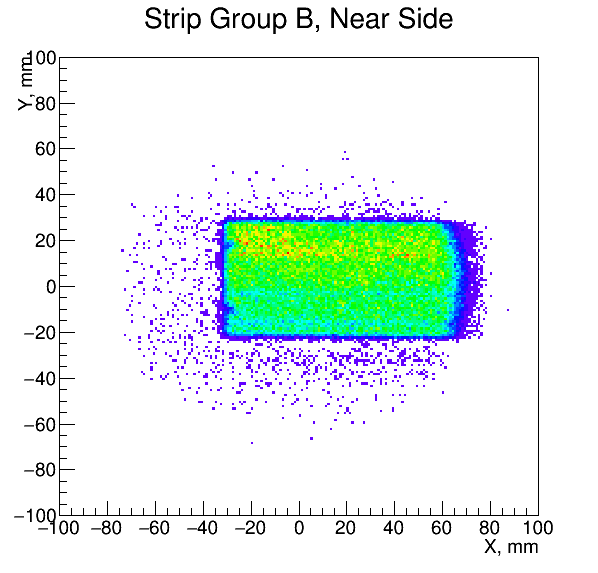} &
\includegraphics[width=.32\textwidth, trim= 5mm 0mm 16mm .08\textwidth, clip]{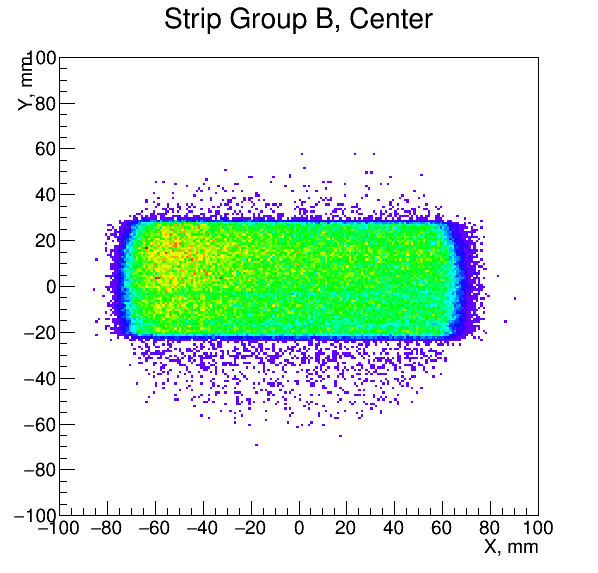} &
\includegraphics[width=.32\textwidth, trim= 5mm 0mm 16mm .08\textwidth, clip]{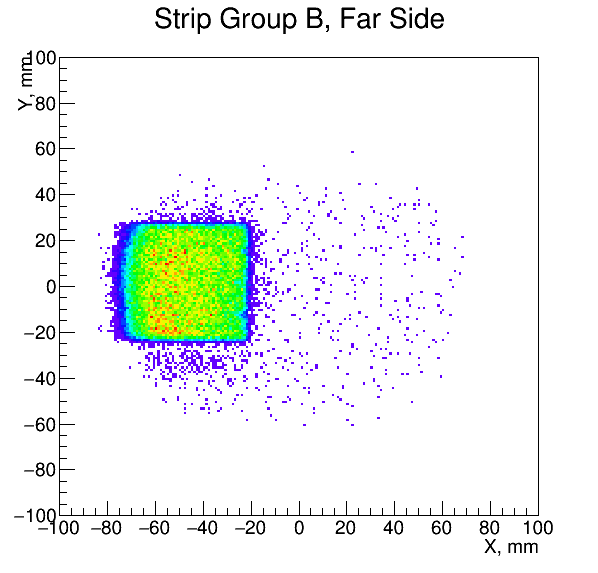} \\
a) & b) & c) \\
\end{tabular}
\caption{\label{fig:expos} Coordinates of the beam tracks with hits in group D strips for the rightmost (a), centered (b) and 
the leftmost (c) positions of the lightproof box.}
\end{figure}

The test setup is sketched in figure~\ref{fig:setup}. The lightproof box with the strips arranged in two groups U0-U2 and D0-D3 was located 
on a table inside a set of three proportional chambers PC1-PC3~\cite{prop}. Each chamber had sensitive wires in both horizontal H and 
vertical V directions with the pitch of 1~mm, while the tracking efficiency was above 99\%. The beam of negative pions with 
the momentum 730~MeV/c was defocused to form the largest possible spot in the area of the strip location and to reduce the angular 
divergence. The beam intensity was limited to about $10^4$~s$^{-1}$. The iron collimator was intended to eliminate the beam halo but also 
limited the spot size to $\sim$150~mm in the horizontal direction. The box could be moved along the alignment rail in the 
horizontal direction, perpendicular to the beam axis, so that different parts of the strip assembly can be exposed to the beam. 
Typically the movement step was 100~mm. The vertical position of the table could be adjusted so that either group U or group D
appeared in the beam center. Figure~\ref{fig:expos} illustrates the illumination of a strip by the beam spot in three positions
of the lightproof box.

The beam trigger was based on the capability of the PC electronics to produce a signal of majority coincidence on more 
than 4 hits in 6 sensitive planes. The trigger was then propagated to the data acquisition system similar to that used
in the DANSS experiment~\cite{wfd}. Waveform digitizers of the DAQ provide both amplitude and time information from the 
SiPM pulses. The calibration of single photo-electron signals was based on thermal noise spectra
from the SiPMs. This amplitude calibration was monitored during the whole data taking period for all SiPMs individually,
together with their cross-talk values. For the purposes of this analysis it is natural to evaluate all amplitude 
parameters in terms of photo-electrons, so such a calibration procedure appears quite adequate.

\begin{figure}[htbp]
\centering
\begin{tabular}{ccc}
\includegraphics[width=.32\textwidth, trim= 3mm 0mm 0mm .11\textwidth, clip]{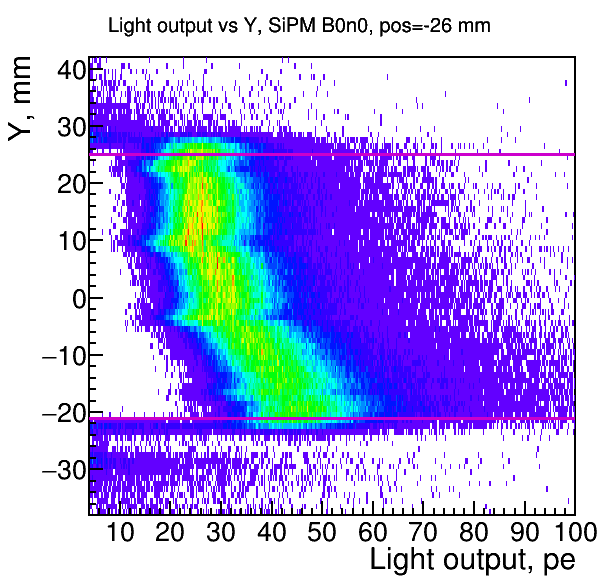} &
\includegraphics[width=.32\textwidth, trim= 3mm 0mm 0mm .11\textwidth, clip]{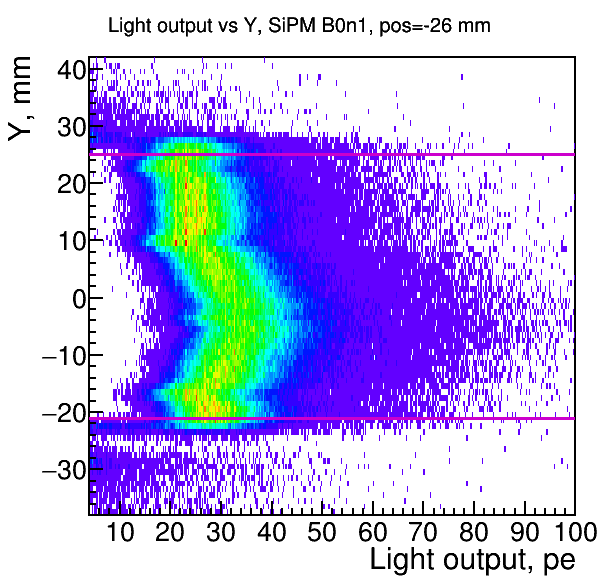} &
\includegraphics[width=.32\textwidth, trim= 3mm 0mm 0mm .11\textwidth, clip]{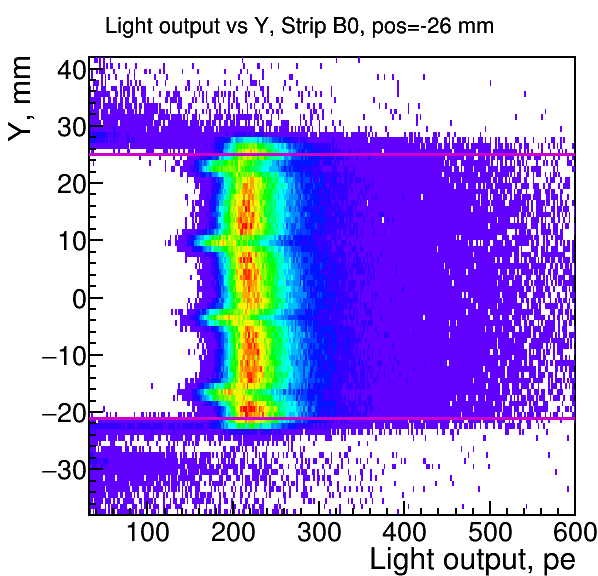} \\
\includegraphics[width=.32\textwidth, trim= 3mm 0mm 0mm .11\textwidth, clip]{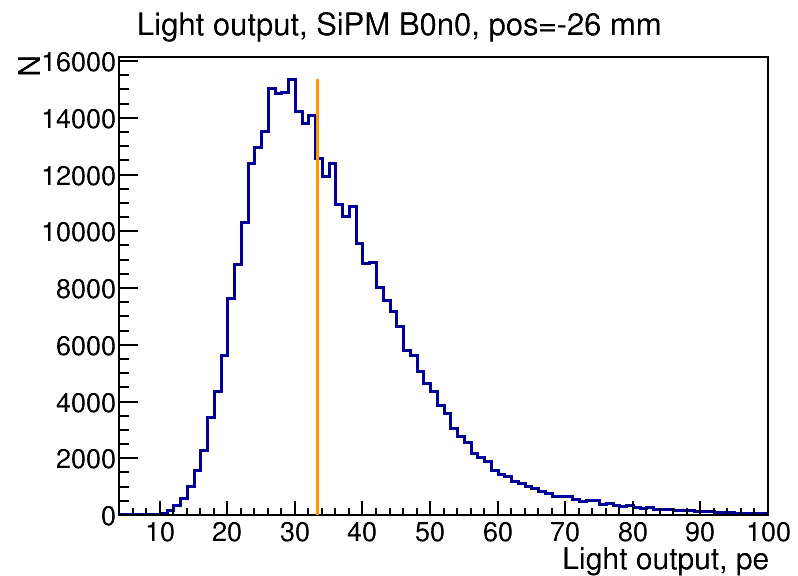} &
\includegraphics[width=.32\textwidth, trim= 3mm 0mm 0mm .11\textwidth, clip]{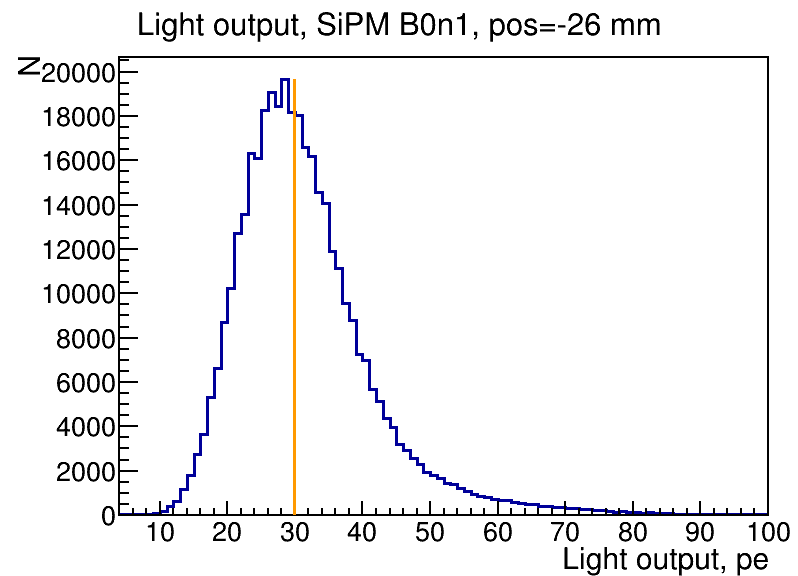} &
\includegraphics[width=.32\textwidth, trim= 3mm 0mm 0mm .11\textwidth, clip]{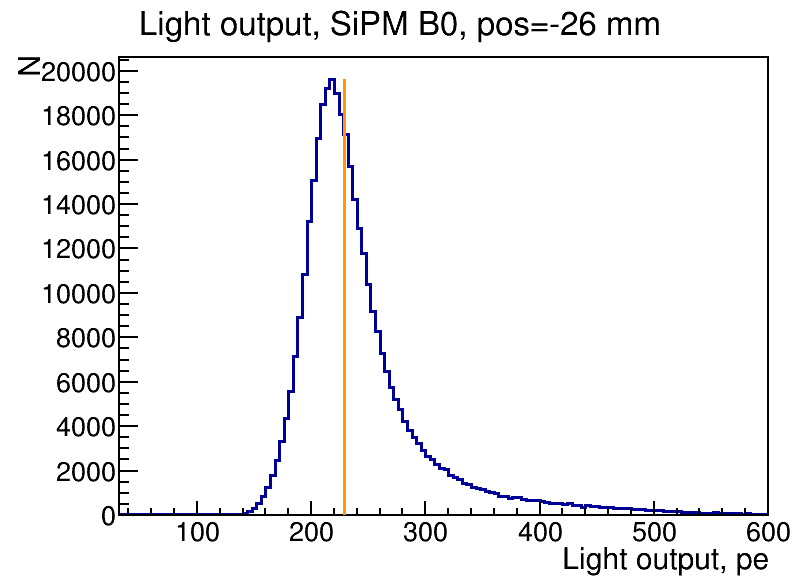} \\
a) & b) & c) \\
\end{tabular}
\caption{\label{fig:eloss} Typical light yield distributions for a side fiber (a), central fiber (b) and 
a sum of all fibers (c) in the central position of the strip assembly.}
\end{figure}

The light collection distributions are very much different in shape for different fibers or their combinations.
In the top row of figure~\ref{fig:eloss} each horizontal band represents the light yield spectrum for a certain
transverse coordinate of the strip. The bottom row shows the corresponding projections from all coordinates (except
2 millimeters on both sides, as indicated by purple lines). In case of significant spectrum shape variations a median value 
of such projection distributions (shown in orange) is chosen as a reasonable measure of the amount of light.

\section{Spatial properties}

\subsection{Longitudinal response profile}

\begin{figure}[htbp]
\centering
\begin{tabular}{cc}
\includegraphics[width=.48\textwidth, trim= 0mm 0mm 0mm .11\textwidth, clip]{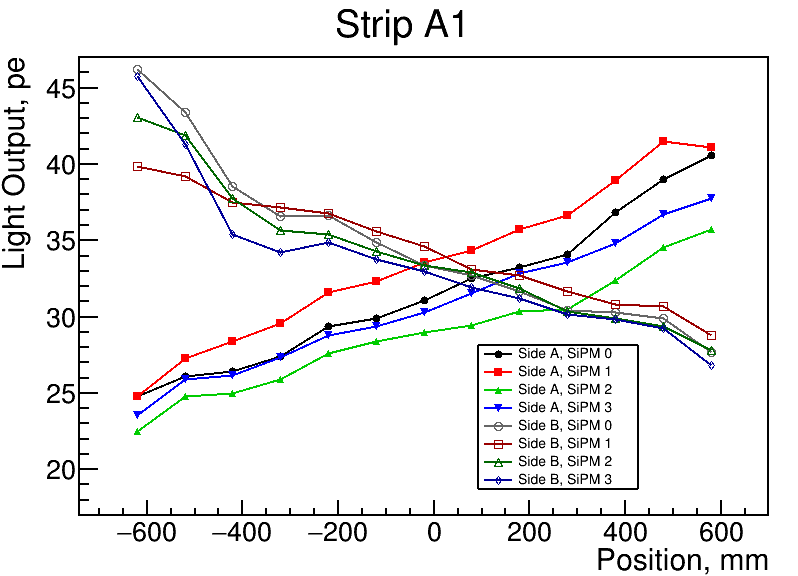} &
\includegraphics[width=.48\textwidth, trim= 0mm 0mm 0mm .11\textwidth, clip]{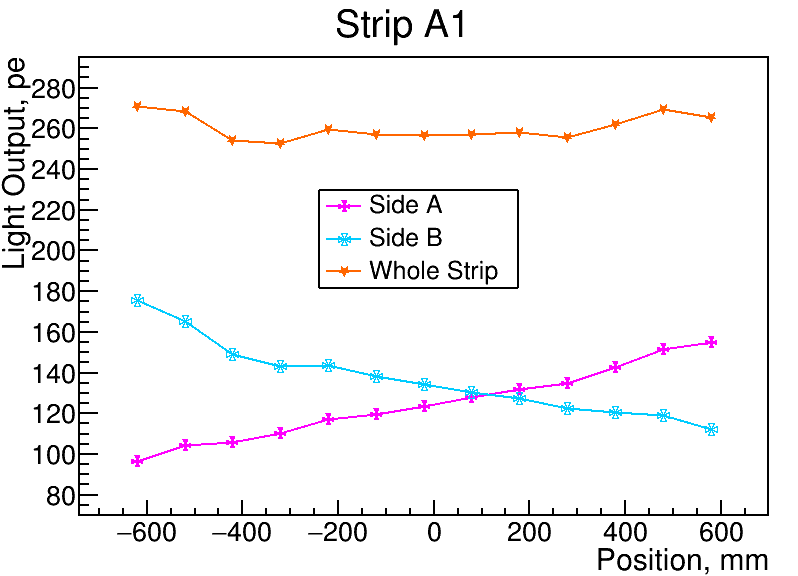} \\
a) & b) \\
\end{tabular}
\caption{\label{fig:lprof} Longitudinal light yield profiles for individual fibers (a),  for the sums at each side 
and for the whole strip (b).}
\end{figure}

The light yield variation along the strtip was measured by changing the position of the lightproof box with 
the strip assembly along the alignment rail, perpendicular to the beam axis. In each position distributions
similar to those in the bottom row of figure~\ref{fig:eloss} were used to determine the median value. Typical 
profiles for individual fibers are presented in figure~\ref{fig:lprof}a. The main reason for the changes in
the light collection along the strip is the attenuation in the fiber. This effect is clearly seen as a light yield
increase when the beam is closer to the photodetector side of the strip. A rough estimate of the attenuation
length gives the value of about 2.8~m. An asymmetric behavior of the two sides may be explained by imperfect
coupling between the fibers and the SiPMs at the side A, possibly because of the misalignment of the printed
circuit board. 

The longitudinal light yield profiles can be measured for each individual strip when deployed as part of a full detector 
using the tracks from the cosmic muons. Once measured, they can be accurately accounted for, given the coordinate 
along the strip is known from the time measurements and/or from the adjacent perpendicular strip hit. At the same
time, a sum of signals from all fibers already gives a reasonably good response (see figure~\ref{fig:lprof}b)
despite of the asymmetric character of individual sides. The profile is fairly flat and only changes
$\pm$3\% (min-max) along the whole strip.

The total light output can be estimated as 75 photo-electrons per MeV in the strip center, while at the
strip sides this value reaches 80~p.e./MeV.

\subsection{Transverse response profile}

The transverse response profiles represent the median values of distributions in 1~mm horizontal bands of the plots in the
top row of figure~\ref{fig:eloss} or similar. The spatial resolution of the PC tracking system is close to this
1~mm value (see edges of the strip images in figure~\ref{fig:expos}). The profiles for individual fibers show
a significant dependence on the distance from the fiber: the collection efficiency drops more than 1.5~times when
the distance becomes larger than the fiber spacing. The groove positions are well seen as small dips at the
corresponding locations. Meanwhile, the total sum for the whole strip and the sums of the four fibers read 
from each side show fairly flat profiles, except for the locations of the grooves, where the decrease reaches~7\%. 
One should expect the perpendicular tracks to be shorter by 20\% at the groove positions, yet the
groove images are blurred by the tracking resolution and thus appear more shallow. 
Also the light collection efficiency increases near the grooves.

\begin{figure}[htbp]
\centering
\begin{tabular}{cc}
\includegraphics[width=.48\textwidth, trim= 0mm 0mm 0mm .11\textwidth, clip]{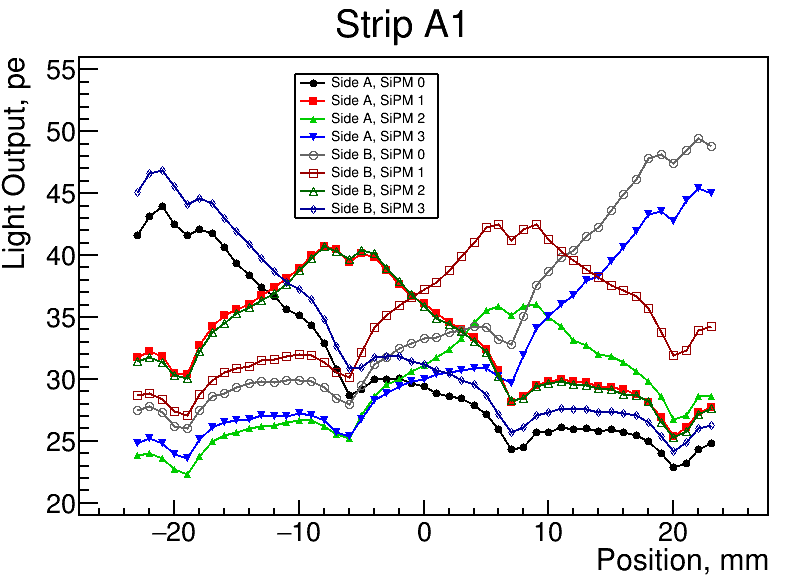} &
\includegraphics[width=.48\textwidth, trim= 0mm 0mm 0mm .11\textwidth, clip]{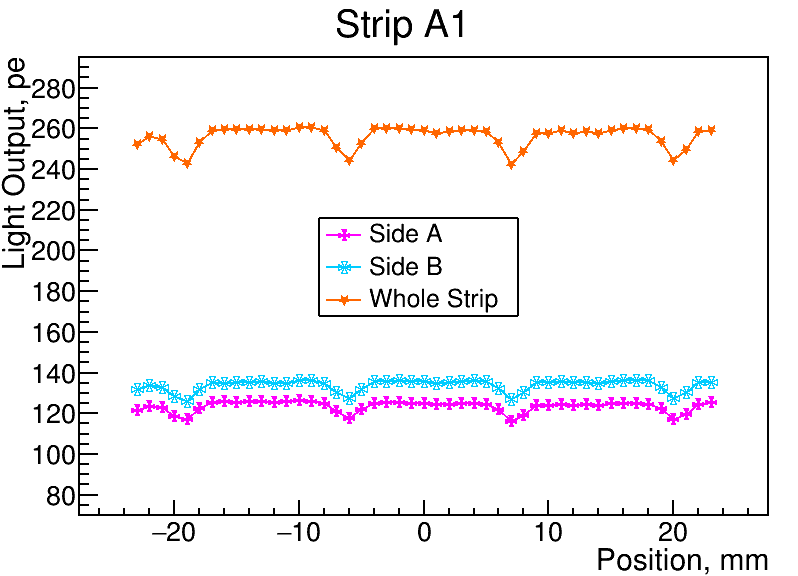} \\
a) & b) \\
\end{tabular}
\caption{\label{fig:tprof} Transverse light yield profiles for individual fibers (a),  for the sums at each side 
and for the whole strip (b). The strip assembly is in the central position.}
\end{figure}

The transverse light yield profiles can not be measured \emph{in situ} during the data taking by the full detector.
The spatial resolution of the muon tracks is insufficient to allow reasonable profile measurement.
Even less is known about the location of the energy deposit for a certain event of interest, so no correction
can be introduced if even the profile is measured. Thus the effect of the transverse profile can only be
estimated as a contribution to the energy resolution. In the case of the whole strip profile in figure~\ref{fig:tprof}b
the r.m.s. deviation is about 2\% and this number corresponds to the tracks perpendicular to the strip plane. 
For the inclined tracks it will be even smaller and hence negligible in comparison to the contribution from the
limited photo-statistics.

\section{Time measurements}

The waveform digitizers used for the data acquisition are sensitive to the system trigger on the
positive transitions of the 125~MHz clocking frequency. Thus the absolute signal time is only known
with the corresponding uncertainty. At the same time, the time between signals corresponding to the same
trigger can be determined with much better accuracy. In this analysis a very simple algorithm resembling
a constant fraction discriminator was applied to find the signal arrival time. The values of two 
sequential points on the waveform that lay below and above the half amplitude level were linearly 
interpolated to this level and the crossing point defined the signal timing. The longitudinal
coordinate of a hit can be determined using the time difference between the signals from the 
opposite sides of the strip. For the best estimate the timing information from 4 fibers read out
at each side was averaged and then the time difference between the sides was calculated. This procedure
preserves the compensation for the common trigger time as it is equivalent to four independent time
difference operations. 

\begin{figure}[htbp]
\centering
\begin{tabular}{cc}
\includegraphics[width=.48\textwidth, trim= 0mm 0mm 0mm .11\textwidth, clip]{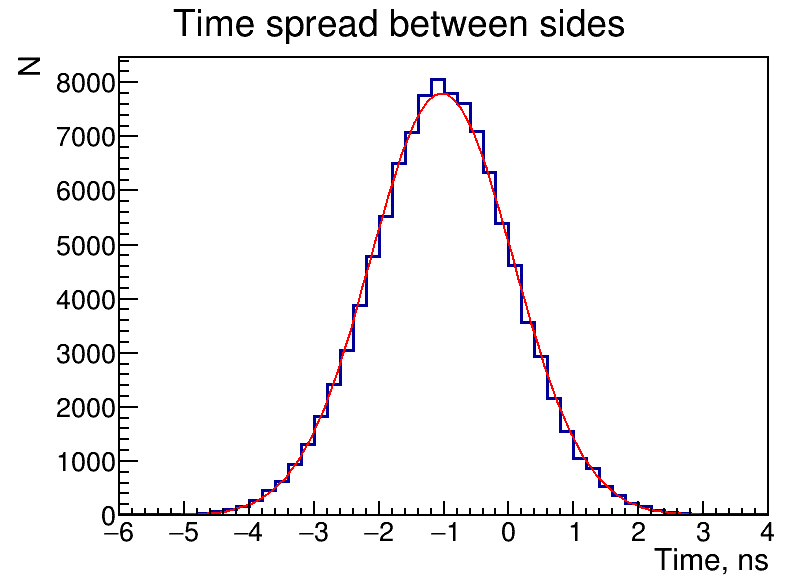} &
\includegraphics[width=.48\textwidth, trim= 0mm 0mm 0mm .11\textwidth, clip]{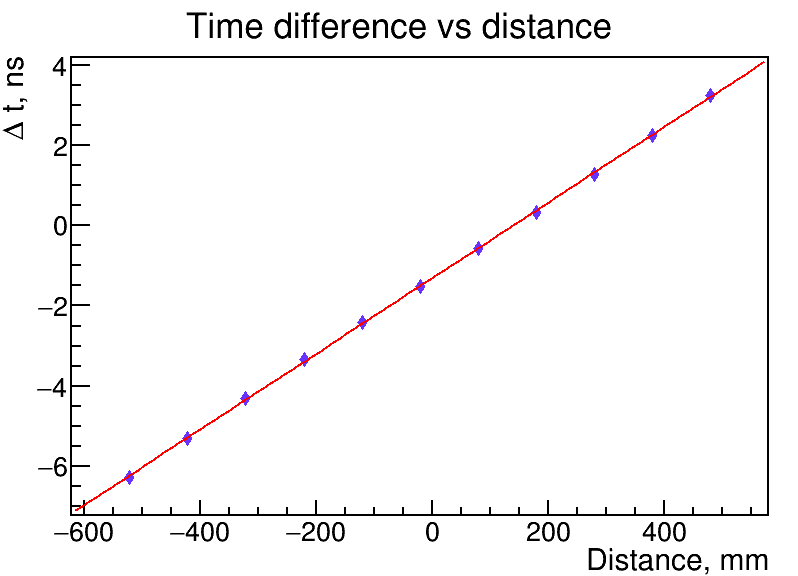} \\
a) & b) \\
\end{tabular}
\caption{\label{fig:tdiff} Time difference distribution (a) between averages from strip sides and its Gaussian fit
with $\sigma=1.1$~ns; dependence of this time difference (b) on the strip assembly position, the slope of the
linear fit is 9.2~ns/m.}
\end{figure}

Even such a simple procedure gives a reasonable timing resolution of 1.1~ns for a MIP crossing the strip perpendicularly
close to its center (see figure~\ref{fig:tdiff}a). To minimize the influence of the coordinate spread on the distribution
width the illuminated area was limited to $\pm20$~mm from the beam center. The propagation delay in the fibers can be 
measured by considering the dependence of the time difference on the longitudinal hit coordinate. Such dependence
is presented in figure~\ref{fig:tdiff}b. The measured slope results in the propagation delay value of 4.6~ns/m
which is in tension with the refraction index n=1.59 of the polystyrene
core of the fiber. This is probably due to the not fully compensated time walk natural for such a simple algorithm.
Yet the method gives quite a linear way of the longitudinal hit position determination and the estimate
for the coordinate resolution is about 12~cm. A more complicated way of the time measurement, for example fitting of the
waveform with a predefined function, may further improve the timing resolution and eliminate the remaining time walk.

\section{Conclusions}

New polystyrene scintillator strips with WLS readout were designed and tested for the DANSS detector upgrade. 
Several improvement ideas resulted in a more than 2 times light yield increase compared to the current detector version.
Optimization of the fiber quantity and of the groove positions led to much better transverse light collection homogeneity, 
resulting in a 2\% r.m.s. spread compared to 8\% in the previous design. A two-sided readout provides the longitudinal coordinate measurement
through the arrival time difference even for the isolated hits. Thus the longitudinal light yield profile can always be accurately
accounted for, though the profile itself is fairly flat even without this correction. The test results are promising
and give hope that the upgraded detector with the strips of the new design will have the aimed energy resolution of 12\% at 1~MeV.

\acknowledgments

The authors are grateful to the staff of the PNPI synchrocyclotron and to the team of the PNPI Meson Physics Laboratory
for providing an excellent pion beam and comfortable test conditions. 

This work is supported by the Ministry of Science and Higher Education of the Russian Federation under the
Contract No. 075-15-2020-778.

\end{document}